\documentclass[%
aps,
prl,
amsmath,amssymb,
superscriptaddress,
reprint,%
nobibnotes
]{revtex4-2}

\usepackage{graphicx}% Include figure files
\usepackage{dcolumn}% Align table columns on decimal point
\usepackage{bm}% bold math
\usepackage[utf8]{inputenc}
\usepackage[T1]{fontenc}
\usepackage{mathptmx}
\usepackage{etoolbox}

\usepackage[dvipsnames]{xcolor}
\usepackage[normalem]{ulem}

\makeatother
\renewcommand{\selectlanguage}[1]{}
\begin{document}
	
	\preprint{AIP/123-QED}
	
	\title[A low-power microstructured atomic oven for alkaline-earth-like elements]{A low-power microstructured atomic oven for alkaline-earth-like elements}
	% Force line breaks with \\
	\author{J. Pick}
	\email{julian.pick@dlr.de}
	\affiliation{Deutsches Zentrum für Luft- und Raumfahrt e.V., Institut für Satellitengeodäsie und Inertialsensorik, Callinstraße 30b, 30167 Hannover, Germany}
	\author{J. Voß}
	\author{S. Hirt}
	\affiliation{LPKF Laser \& Electronics SE, Osteriede 7, 30827 Garbsen, Germany}
	\author{J. Kruse}
	\affiliation{Deutsches Zentrum für Luft- und Raumfahrt e.V., Institut für Satellitengeodäsie und Inertialsensorik, Callinstraße 30b, 30167 Hannover, Germany}
	\author{T. Leopold}
	\affiliation{Deutsches Zentrum für Luft- und Raumfahrt e.V., Institut für Satellitengeodäsie und Inertialsensorik, Callinstraße 30b, 30167 Hannover, Germany}
	\affiliation{LPKF Laser \& Electronics SE, Osteriede 7, 30827 Garbsen, Germany}
	\author{R. Schwarz}
	\author{C. Klempt}
	\affiliation{Deutsches Zentrum für Luft- und Raumfahrt e.V., Institut für Satellitengeodäsie und Inertialsensorik, Callinstraße 30b, 30167 Hannover, Germany}

	\date{\today}
	
	\begin{abstract}
    Alkaline-earth-like elements play pivotal roles in advanced quantum sensing technologies, notably optical clocks, with unprecedented precision achieved in recent years. Despite remarkable progress, current optical lattice clocks still face challenges in meeting the demanding size, weight, and power consumption constraints essential for space applications. Conventional atom sources, such as ovens or dispensers, require substantial heating power, making up a significant fraction of the system's overall power consumption. Addressing this challenge, we present a novel microstructured atomic oven based on fused silica, designed for miniaturization and low-power operation. We characterize the oven by loading a magneto-optical trap  with Yb evaporated from the oven and demonstrate operation with a loading rate above $10^8\,\mathrm{atoms}/\mathrm{s}$ for heating powers below $250\,$mW.
	\end{abstract}
	
	\maketitle

	\section{\label{sec:introduction}Introduction}
	
	Alkaline-earth-like elements are employed in state-of-the-art quantum sensors, most prominently optical clocks \cite{takamoto_optical_2005, ludlow_optical_2015}, and in quantum computing with neutral atoms \cite{jenkins_ytterbium_2022, ma_universal_2022} or trapped ions \cite{bruzewicz_trapped_2019}. Rapid technological advancement enabled the operation of optical clocks with fractional frequency uncertainties in the low $10^{-18}$ range \cite{ushijima_cryogenic_2015, mcgrew_atomic_2018, bothwell_jila_2019, brewer_al_2019, huang_liquid_2022, tofful_171_2024} 
	and the development of transportable systems aiming to field deployment \cite{koller_transportable_2017,  fasano_transportable_2021, ohmae_transportable_2021, huang_geopotential_2020} or even reaching out into space \cite{origlia_towards_2018, chen_development_2024} where numerous applications for optical lattice clocks are proposed \cite{derevianko_fundamental_2022,schkolnik_optical_2023}.
	
	Today's maturity of optical lattice clocks is still insufficient to meet the stringent requirements of a space flight in terms of size, weight and power consumption (SWaP). Field-deployable systems, either on ground or in space, therefore require significant miniaturization of key components.
	For alkaline-earth-like elements, conventional atom sources are based on an oven \cite{schioppo_compact_2012, song_cost_2016, wodey_robust_2021} or dispensers \cite{bridge_vapor_2009, dorscher_creation_2013, kwon_jet_2023, nomura_direct_2023} that require heating to temperatures up to $400-500^\circ\,$C in order to provide a sufficient flux of atoms, where typical heating powers of several tens of watts are required. The evaporated atoms are then generally trapped and cooled in a magneto-optical trap (MOT) as a first step in the preparation of ultra-cold atoms. In transportable experiments, the required heating power of the source can be a significant portion of the system's overall power consumption \cite{chen_development_2024}.
	
	A mitigating technique in the reduction of thermal losses due to convection is by using in-vacuum heating \cite{schioppo_compact_2012}, leaving thermal radiation and -conduction as the remaining loss processes. Still the required heating power remains in the order of tens of watts.
	Progress towards the reduction of thermal conduction has been made with a chip-size atomic oven based on silicon, which is suspended by narrow beams \cite{schwindt_highly_2016}. The achievable reduction of thermal conduction however was limited due to the decreasing mechanical robustness with reduced beam thickness.
	
	In this article, we present a miniaturized monolithic in-vacuum atomic oven based on microstructured fused silica. Compared to silicon, fused silica offers a reduction of the thermal conductance by two orders of magnitude. The oven was manufactured by Laser Induced Deep Etching (LIDE).
	The monolithic design holds a spring-mounted reservoir that contains the atoms and is heated directly. A laser-structured electrical circuit allows for electrical heating as well as heating by optical absorption of light. While evaporation of alkaline-earth-like atoms by optical illumination has already been demonstrated \cite{kock_laser_2016, yasuda_laser_2017}, our design relaxes the optical power requirement and enables a continuous operation without the need for a regular realignment of the optical heating beam. 
	The geometry of the mounting springs offers optimal thermal insulation of the reservoir from the mounting structure as well as mechanical stability with respect to differential thermal expansion along the springs themselves. Thus, thermal radiation remains as the only significant loss process at high temperatures, enabling an oven operation at low power consumption.
	Compared to conventional atomic beam ovens, our approach drastically reduces size and weight. Therefore it can be placed directly near the structure that generates the MOT. 
	Although our atomic reservoir is likewise smaller compared to classical oven designs, we expect a sufficient lifetime on the order of multiple years since the atomic flux into the MOT volume is comparable at much lower oven temperatures i.e. atomic evaporation rates. For both heating options we demonstrate a power consumption far below $1\,$W for generating a fast loading MOT with more than $10^8\,\mathrm{atoms}/{\mathrm{s}}$.
	
	We compare both heating mechanisms the oven features and characterize its atom evaporation properties by measuring the MOT loading dynamics.
    Our oven is specifically well suited to operate modern, compact, single-beam MOT structures \cite{bowden_pyramid_2019, sitaram_confinement_2020, bondza_two_2022, bondza_achromatic_2024} which further miniaturize a crucial component of modern quantum sensors, paving the way towards space-borne applications.
    
	\section{\label{sec:design}Oven design}
	\begin{figure*}
		\centering
		\includegraphics[width = 0.9\textwidth]{"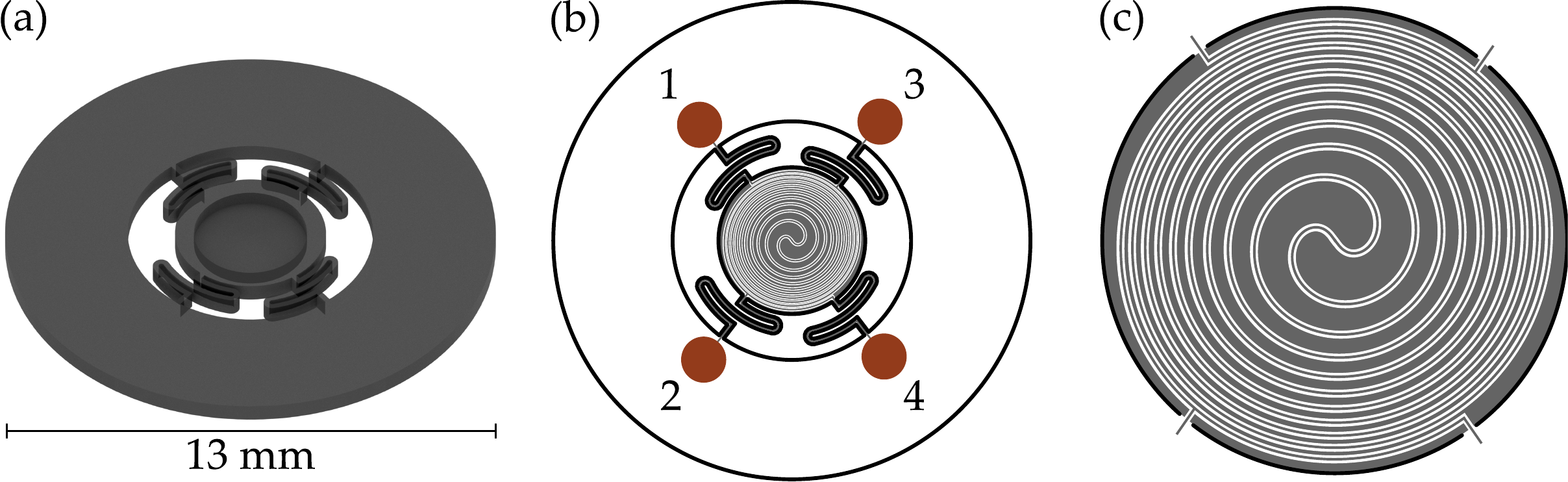"}
		\caption{\label{fig:microoven_yb}Design of the monolithic microstructured fused silica oven. (a) Rendered image of the front side showing the spring mounted reservoir in the center. (b) Sketch of the back side. The numbered brown circles indicate copper contacting pads and gray lines represent the platinum layer. (c) Magnified view of the back side of the heating plate. Here the high platinum filling factor supports light absorption for improved optical heating from the back side. Not shown are the mounting springs and the outside part of the chip.}
	\end{figure*}
	The atomic oven is depicted in Fig.~\ref{fig:microoven_yb}.
	It is based on a circular fused silica chip that is manufactured in larger numbers from a 6" fused silica wafer with $500\,$µm initial thickness. An outer support structure of $13\,$mm diameter holds the central part via four double folded mounting springs, each with a cross-section of $150\,$µm$\times 450\,$µm and a length of $6\,$mm. The central part of the oven is a heating plate with $4\,$mm outer diameter and $3\,$mm wide recess. It can be filled with alkaline-earth-like atoms and is the only part of the chip that is heated in order to evaporate the atoms.
	
	The suspension of the heating plate in the chip using springs is a key feature in the thermal isolation of the heating plate towards the mounting structure. It minimizes heat conductance from the heating plate to the outside part of the chip, due to its favorable aspect ratio between cross section and length. Furthermore it ensures mechanical stability with respect to thermal expansion along the spring, which would introduce mechanical stress and out-of-plane bending in a straight connection due to the expected large temperature gradient between the mounting structure and the heating plate.
	The chips were fabricated using the two-step LIDE technology developed by LPKF \cite{ostholt_high_2014}. In the first step, the glass is laser-modified, and in the second step chemically etched with a hydrofluoric acid-based solution. The laser-modified areas etch at a significantly higher rate than the unmodified ones, enabling the formation of high aspect ratio structures with a precisely defined geometry without introducing defects into the glass. The defect-free manufacturing process provides the mechanical stability needed for this application. During the etching process, the overall substrate thickness is reduced to about $450\,$µm which was taken into account in the design of the heating plate. 
	
	The back side of the chip was metalized in a physical vapor deposition (PVD) process to enable electrical heating with a $300\,$nm platinum layer on a $10\,$nm titanium adhesion layer. 
	A heating structure in the shape of a Fermat's spiral was manufactured by removing a $15\,$µm wide stripe of the metal layer via laser scribing, creating an electrical insulation between neighboring windings. The remaining conductive path along the spiral is approximately $65\,$µm wide with a total length of about $84\,$mm. 
	Each ending of the spiral is split into two parallel paths, where the conductor length difference for two parallel connections is about $2.5\,$mm. Each path is terminated by a copper contacting pad. Two of the copper pads (1 and 4 in Fig.~\ref{fig:microoven_yb}) are connected to an electric power source for electrical heating while the other two (contacting pads 2 and 3) are used for measuring the voltage drop along the spiral enabling a four-point resistance measurement, to determine the spiral's temperature. 
	Apart from the electrical heating, the reservoir can also be heated by optical illumination. For that, the Pt filling between the lanes, which was not removed, maximizes the area that effectively absorbs the heating light.
	The bottom of the reservoir separates the heated Pt layer from the atoms. In order to ensure sufficient heat transfer without compromising mechanical stability, it has a thickness of $30\,$µm.
	
\begin{figure}
		\centering
		\includegraphics[width = 0.45\textwidth]{"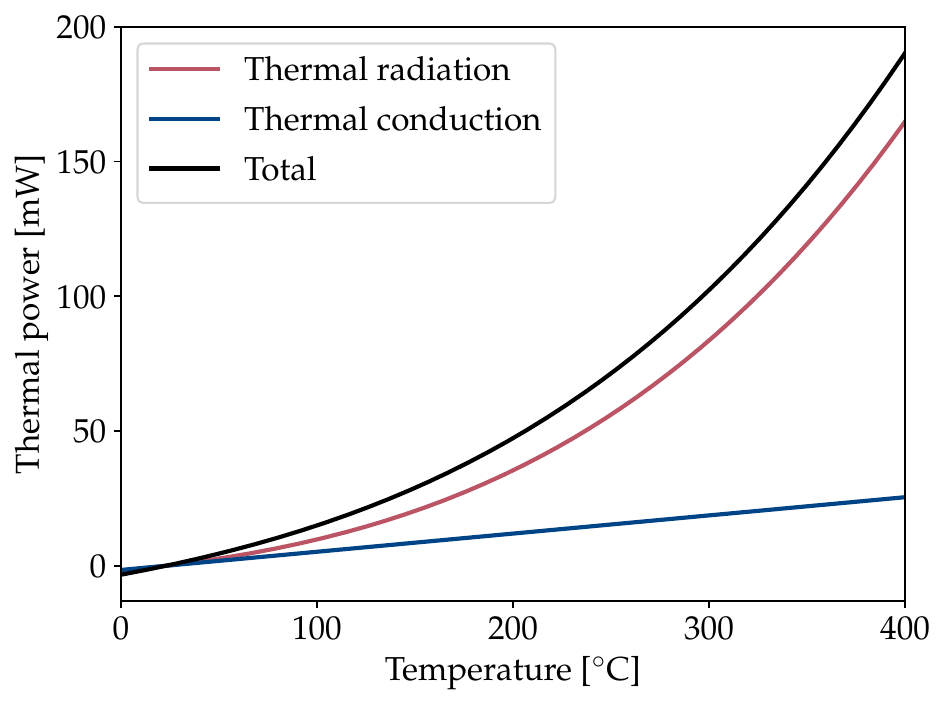"}
		\caption{\label{fig:thermal_power}Estimated thermal power loss as a function of temperature of the heating plate.}
	\end{figure}

	When the oven is operated in-vacuum, the power consumption is determined by losses due to thermal radiation and thermal conduction, where thermal radiation is described by the Stefan-Boltzmann law and thermal conduction by Fourier's law. For our oven geometry, we calculate an estimate of the total power consumption as a function of the heating plate temperature, which is shown in Fig.~\ref{fig:thermal_power}. It can be seen that for temperatures above $100^\circ\,$C, the total power loss is dominated by thermal radiation, since heat conduction through the mounting springs is suppressed due to their geometry and the material properties of fused silica. 
	
\begin{figure}
		\centering
		\includegraphics[width = 0.45\textwidth]{"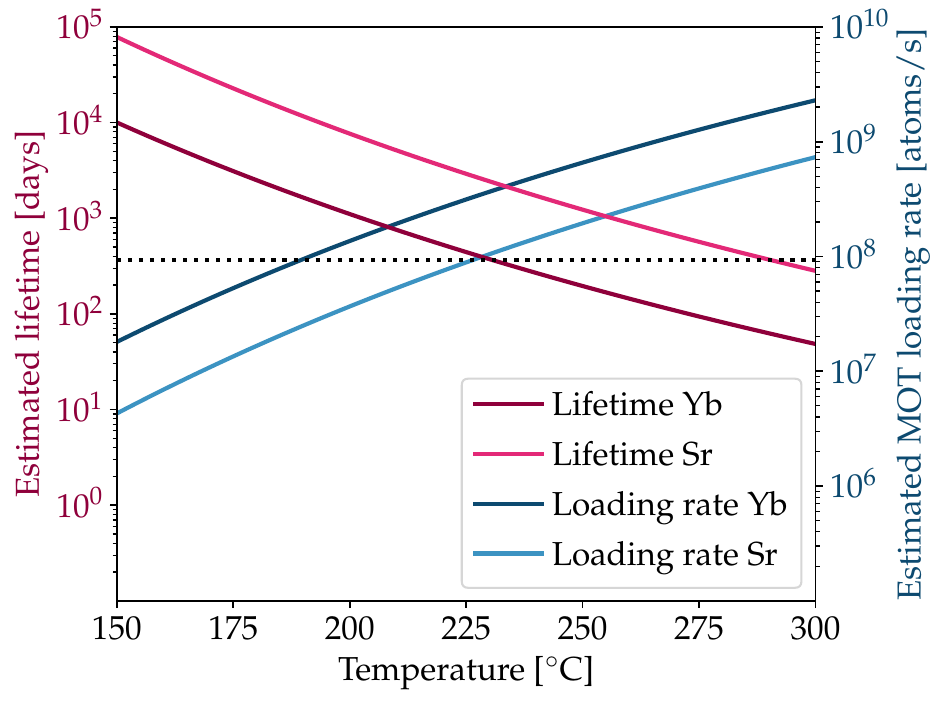"}
		\caption{\label{fig:microoven_lifetime}Estimated oven lifetime and MOT loading rate as a function of temperature for Yb and Sr. The black dotted line indicates a lifetime of one year. MOT loading rates are given for the most abundant isotopes $^{174}$Yb and $^{88}$Sr.}
	\end{figure}

	Due to the comparably small volume of the reservoir of only about $3\,\mathrm{mm}^3$, a critical design parameter is the oven lifetime, which means the duration for how long it can sustain a decent flux of atoms before the reservoir becomes empty. At a given oven temperature the lifetime can be directly calculated from the expected evaporation rate of the specific element that is used.
	We calculated the vapor pressure $p_\mathrm{v}$ using \cite{safarian_vacuum_2013}
	\begin{equation}
	\label{eq_vappress}
	\log(p_\mathrm{v}) = A/T + B \log(T) + C \cdot T + D,
	\end{equation}
	with parameters $A$, $B$, $C$, $D$ taken from literature for Yb and Sr.
	The corresponding evaporation rate at temperature $T$ is given by the Hertz-Knudsen-equation:
	\begin{equation}
	\label{eq_evaprate}
	\dot{n}(T) = \frac {p_\mathrm{v}(T)} {\sqrt{2 \pi \cdot M \cdot R \cdot T}} \cdot N_\mathrm{a},
	\end{equation}
	where $M$ is the molecular mass, $R$ the ideal gas constant and $N_\mathrm{a}$ the Avogadro constant.
	From the specific evaporation rates, we can also estimate the MOT loading rate that can be achieved.  
	Here the small oven size emerges as a key advantage, enabling to place it directly next to the MOT volume and thus gaining a comparably large geometric acceptance angle for evaporated atoms to enter the MOT trapping volume. In the setup described in the following section, the half-opening angle for which atoms can enter the MOT volume is about $22^\circ$. Further we assume a maximum MOT capture velocity of our pyramid MOT of $40\,\mathrm{m}/\mathrm{s}$. From this we estimate MOT loading rates of the most abundant isotopes of the alkaline-earth-like elements $^{174}$Yb and $^{88}$Sr for different temperatures of the evaporated atoms. They are shown in Fig.~\ref{fig:microoven_lifetime}, together with the expected lifetimes of a reservoir with an initial filling of $30\,$mg. It is visible that for both elements, continuous operation with MOT loading rates above $10^8\,\mathrm{atoms}/\mathrm{s}$ for multiple years is possible. Furthermore, the lifetime increases drastically when the oven is operated at a lower temperature yielding more conservative loading rates, or when the oven is not operated continuously. An even higher lifetime could be reached by redesigning the oven with a larger reservoir, at the cost of slower heating dynamics and an increase of the overall size and weight.
	
	The reservoir can be filled by evaporation of atoms from an external source and subsequent deposition on the glass substrate \cite{manginell_in_2012,schwindt_highly_2016}, but for the results described in the following we instead placed a solid block of ytterbium with a weight of approximately $30\,$mg onto the reservoir. A thin layer of indium foil was added between the Yb block and the reservoir, in order to improve the thermal contact to the heating plate and thus reduce the required electrical heating power. 

	\section{Results} \label{sec:results}
\begin{figure}
		\centering
		\includegraphics[width = 0.35\textwidth]{"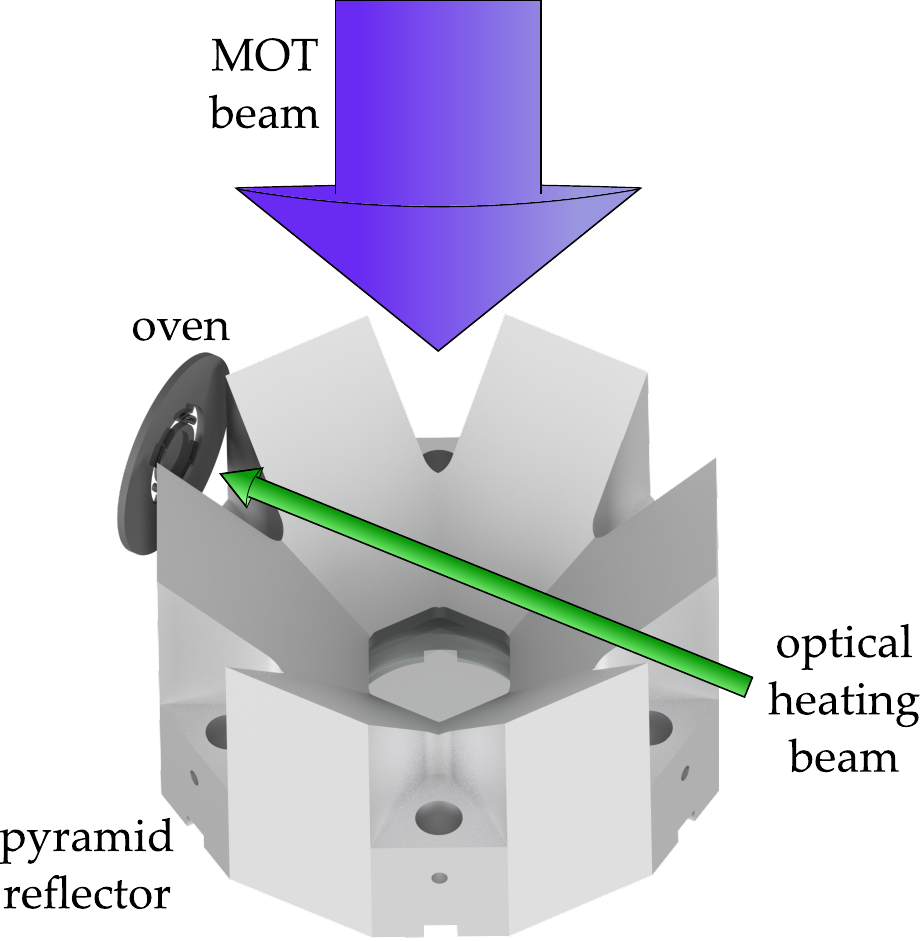"}
		\caption{\label{fig:Oven_in_Vacuum}Experimental setup for the characterization of the oven.}
	\end{figure}

	We test the oven by loading a MOT generated by a pyramid reflector \cite{lee_single_1996} that only requires a single incident laser beam. The reflector is a quasi-monolithic aluminium structure and is described in detail in Ref. \cite{pick_compact_2024b}. It features six angled reflective surfaces to generate beams for radial trapping and a bichromatic waveplate with a highly-reflective coating on its back side for axial trapping. Gaps between the reflective surfaces allow for radial optical access and atom loading. The oven is placed directly at such a gap with a distance of approximately $1\,$mm from the outside of the reflector, as indicated in Fig.~\ref{fig:Oven_in_Vacuum}.
	This enables atoms that are evaporated from the oven under a large solid angle to enter the trap volume. The exact position of the oven and distance to the pyramid can be adjusted with a port aligner.
	
	For electrical heating, we applied a voltage between the connections 1 and 4 and simultaneously measured the voltage drop across connections 2 and 3. We measured a room-temperature resistance of the spiral of $523\,\Omega$. However, it was observed that the resistance does not monotonously increase with increasing heating power, so that it is not possible to determine the temperature from the measured resistance.
	
	The gap in the pyramid opposing the oven is used for coupling in a beam at $556\,$nm for optical heating. The optical heating beam has a diameter of $4\,$mm, so that the reservoir is fully illuminated. With this optical heating setup, the Yb is heated directly by absorbing the $556\,$nm light, instead of the Pt layer, which has a low absorptance at this wavelength. We chose this setup, because direct illumination of the Yb block was easier due to the good optical access through the pyramid. In this case, the oven is not used directly for heating, but still ensures the thermal isolation from the environment for a low power requirement on the heating beam. Direct heating of the Pt layer would require a lower wavelength for efficient optical absorption.
	
    The magnetic quadrupole field that is required for MOT operation is generated by two coils in anti-Helmholtz configuration, which are wound on two water-cooled copper mounts. The magnetic field gradient was $3.3\,\mathrm{mT}/\mathrm{cm}$ in axial direction for all measurements.
    A single laser beam at $399\,$nm is incident on the pyramid reflector and is split and reflected into all required beams for three-dimensional trapping and cooling. We choose a collimated Gaussian beam with $1/e^2$ diameter of $45\,$mm, which is larger than the pyramid's outer diameter of $34\,$mm.
    
\begin{figure}
		\centering
		\includegraphics[width = 0.45\textwidth]{"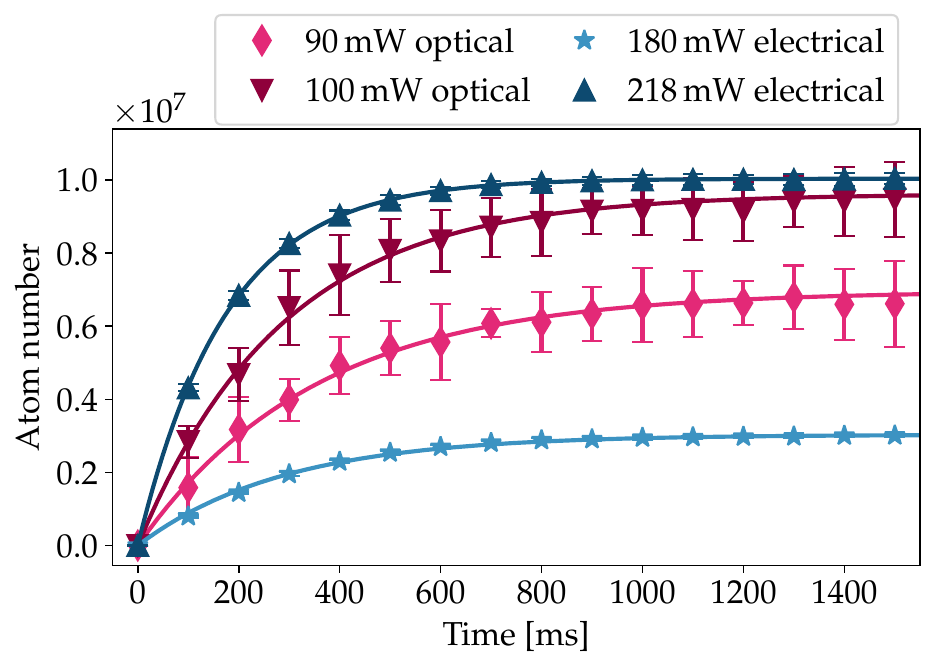"}
		\caption{\label{fig:mot_loading_curves}Exemplary MOT loading curves for different optical and electrical heating powers.}
	\end{figure}

\begin{figure}
		\centering
		\includegraphics[width = 0.45\textwidth]{"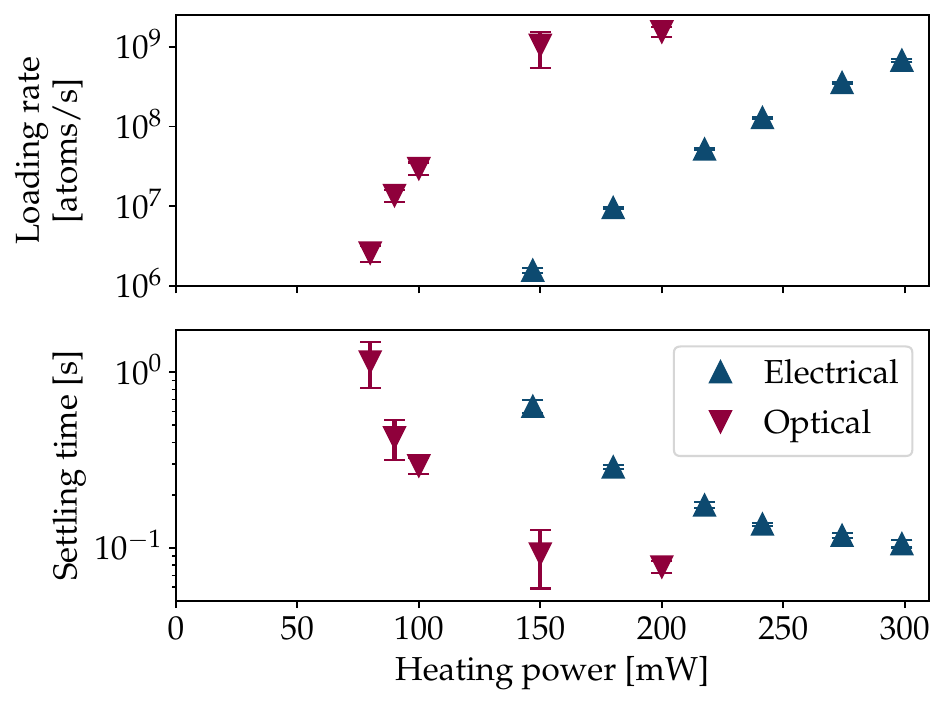"}
		\caption{\label{fig:mot_loading_rates}MOT loading rates and settling times as a function of heating power for both heating mechanisms.}
	\end{figure}

    We load a MOT operating on the broad $^1S_0 \rightarrow \,^{1}P_1$ transition at $399\,$nm. We measured MOT loading curves of the most abundant isotope $^{174}$Yb for different electrical and optical heating powers. The optical power for the MOT at $399\,$nm was $60\,$mW and the MOT detuning was $-32\,$MHz. Exemplary loading curves are shown in Fig.~\ref{fig:mot_loading_curves}. Here, error bars indicate statistical fluctuations and represent one standard deviation. Solid lines are exponential fits, that yield the respective loading rates and settling times. Here, the settling time is equivalent to the inverse of the atomic loss rate in the fully loaded MOT. Both parameters are shown for different heating powers in Fig.~\ref{fig:mot_loading_rates}. It is visible that with both heating mechanisms loading rates above $10^8\,\mathrm{atoms}/\mathrm{s}$ can be achieved with heating powers below $250\,$mW. In order to achieve the same loading rates, the required power for electrical heating is larger than for optical heating. This is due to the limited thermal contact between the Yb block and the reservoir, which is advantageous when the Yb is directly heated by the optical beam and disadvantageous for electrical heating of the reservoir. Nevertheless, electrical heating is potentially the more power-efficient approach, since the generation of laser light requires more electrical power than the optical output power that is available for heating.
    For both heating mechanisms, the MOT lifetime decreases with increasing heating power. This can be explained by collisions of the trapped atoms with fast atoms emerging from the oven, that limit the lifetime and scale with heating power.
	
\begin{figure}
		\centering
		\includegraphics[width = 0.45\textwidth]{"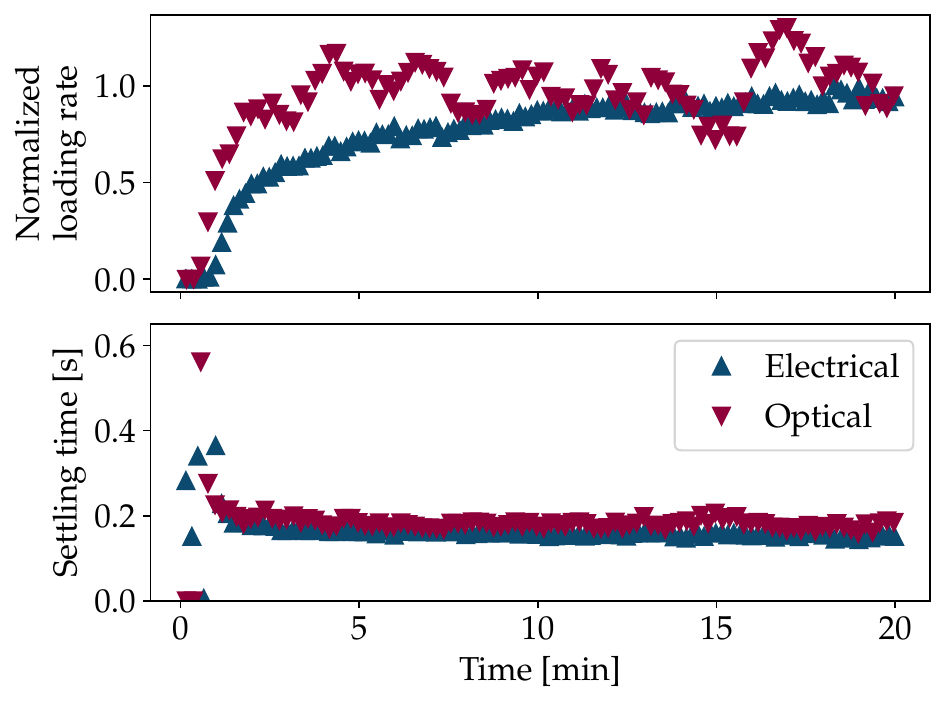"}
		\caption{\label{fig:heating_dynamics}Heating dynamics of the oven with electrical and optical heating. The MOT loading rates for electrical heating are normalized to the maximal value and for the optical heating they are normalized to the mean value in the interval between $5\,$min and $20\,$min.}
	\end{figure}

	We investigated the heating dynamics of the oven by repeatedly loading a MOT and measuring the loading rate. Starting at room temperature, we turned on the electrical or optical heating with powers of $205\,$mW and $100\,$mW respectively and tracked the loading rate over a duration of $20\,$min. The result is shown in Fig.~\ref{fig:heating_dynamics}. For the electrical heating, the loading rate strongly increases within the first $2\,$min and then slowly increases further during the whole measurement. For the optical heating, the loading rate also strongly increases during the first $2\,$min, but then oscillates. This periodic oscillation can be associated to power fluctuations of the optical heating beam, which is why it does not appear for the electrical heating. This means that the electrical heating generates a far more stable flux of atoms compared to the optical heating, when the optical heating power is not actively stabilized.
	
\begin{figure}
		\centering
		\includegraphics[width = 0.45\textwidth]{"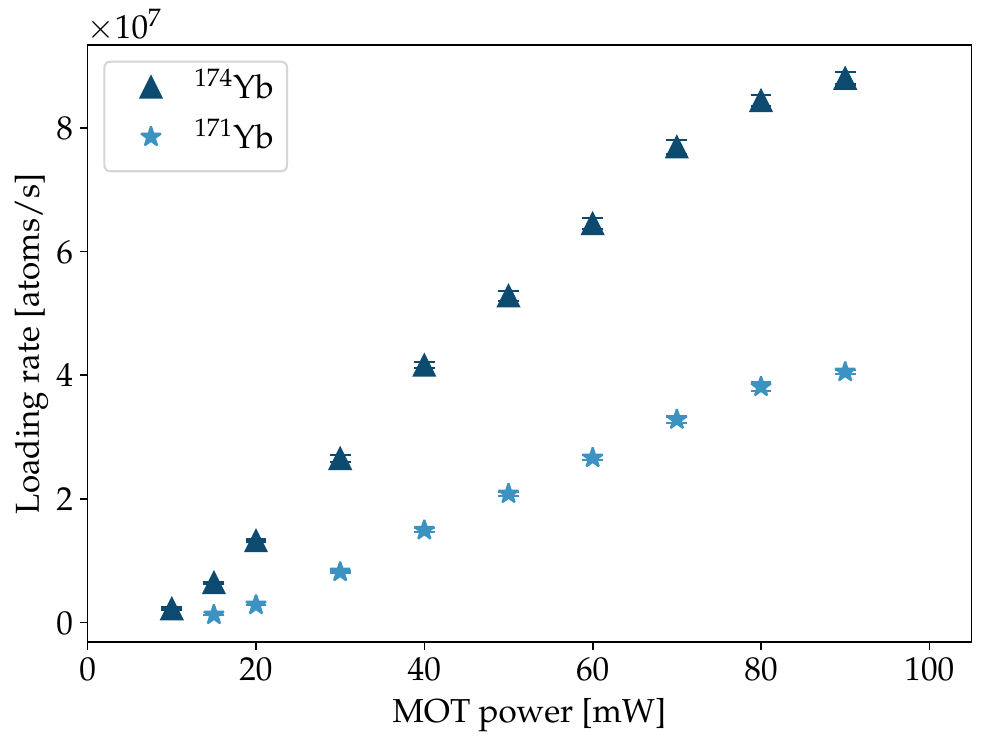"}
		\caption{\label{fig:loading_rate_vs_power_compare_isotopes}MOT loading rate as a function of MOT power at $399\,$nm for both relevant isotopes $^{174}$Yb and $^{171}$Yb. The electrical heating power was $205\,$mW and the MOT detuning was $-32\,$MHz.}
	\end{figure}

	The velocities of the atoms that are evaporated from the oven are described by a thermal distribution. Due to the high temperature that is required for evaporation, it can be assumed that only a small fraction of the atoms is slow enough to be captured by the MOT. It can thus be expected that the MOT capture velocity, which can be enhanced by increasing the MOT power, has a strong influence on the loading rate. For a fixed electrical heating power of $205\,$mW we measured the loading rate as a function of MOT power. Additionally to $^{174}$Yb, the fermionic isotope $^{171}$Yb is of interest, since it is commonly used in optical lattice clocks. We performed the measurement for both isotopes (Fig.~\ref{fig:loading_rate_vs_power_compare_isotopes}). It can be seen that the respective loading rates strongly increase with increasing MOT power. The loading rates for the fermionic isotope are lower due to its lower natural abundance. Nevertheless, it can be seen that a high loading rate of $^{171}$Yb can be achieved with MOT powers that are typically available from commercial laser sources.
	
\begin{figure}
		\centering
		\includegraphics[width = 0.45\textwidth]{"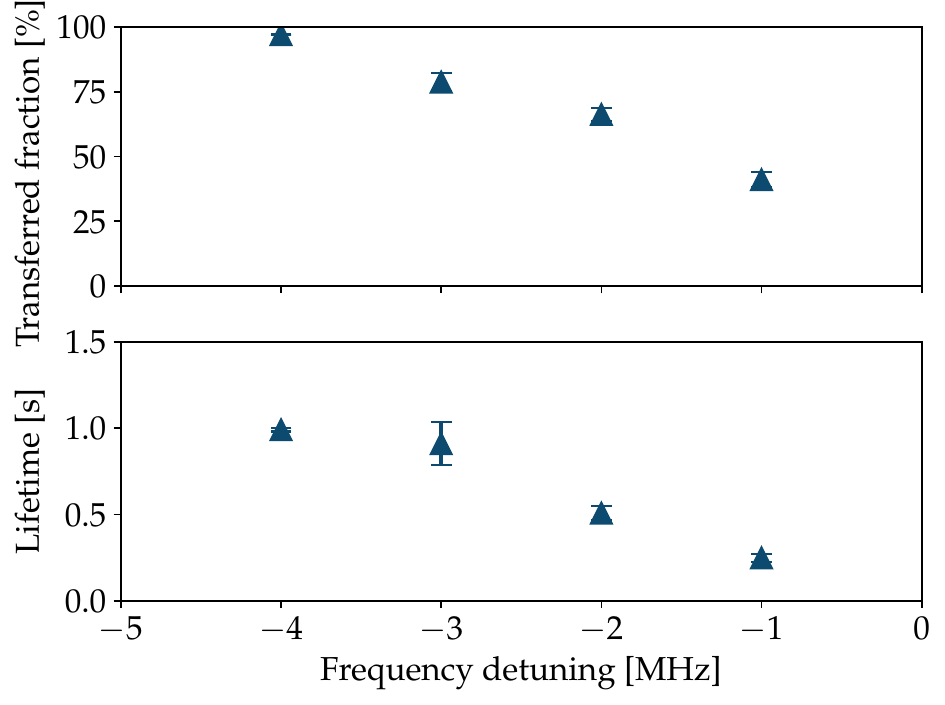"}
		\caption{\label{fig:transfer_efficiency_vs_detuning}Transfer efficiency and lifetime in the second MOT stage as a function of frequency detuning of the $556\,$nm MOT laser.}
	\end{figure}

	When the oven is heated with an electrical power of $205\,$mW, the pressure inside the vacuum chamber rises to $5\times10^{-10}\,$mBar. Although the fast atoms evaporating from the oven can collide with the trapped atoms, it is possible to efficiently transfer the atoms to the second MOT stage operating on the narrow $^1S_0 \rightarrow \,^{3}P_1$ transition at $556\,$nm. The MOT beam at $556\,$nm is incident on the pyramid reflector through the same optical path as the first-stage MOT beam, resulting in the same beam diameter. In order to characterize the second MOT stage, we first load a MOT on the first-stage transition. We then switch on the $556\,$nm light and after a transfer time of $20\,$ms we turn off the $399\,$nm light. After a variable holding time, we transfer the atoms back into the first MOT stage and compare the initial and final atom numbers. This atom number ratio is measured for different holding times in the second MOT stage, so that the inital transfer efficiency and the lifetime can be determined. We repeated this measurement for different detunings of the $556\,$nm laser (Fig.~\ref{fig:transfer_efficiency_vs_detuning}), which had an optical power of $10\,$mW. It is visible that a transfer efficiency above $90\,\%$ is possible, which shows that the oven is ideally suited to serve as an atom source for a compact optical lattice clock.

\begin{figure}
		\centering
		\includegraphics[width = 0.45\textwidth]{"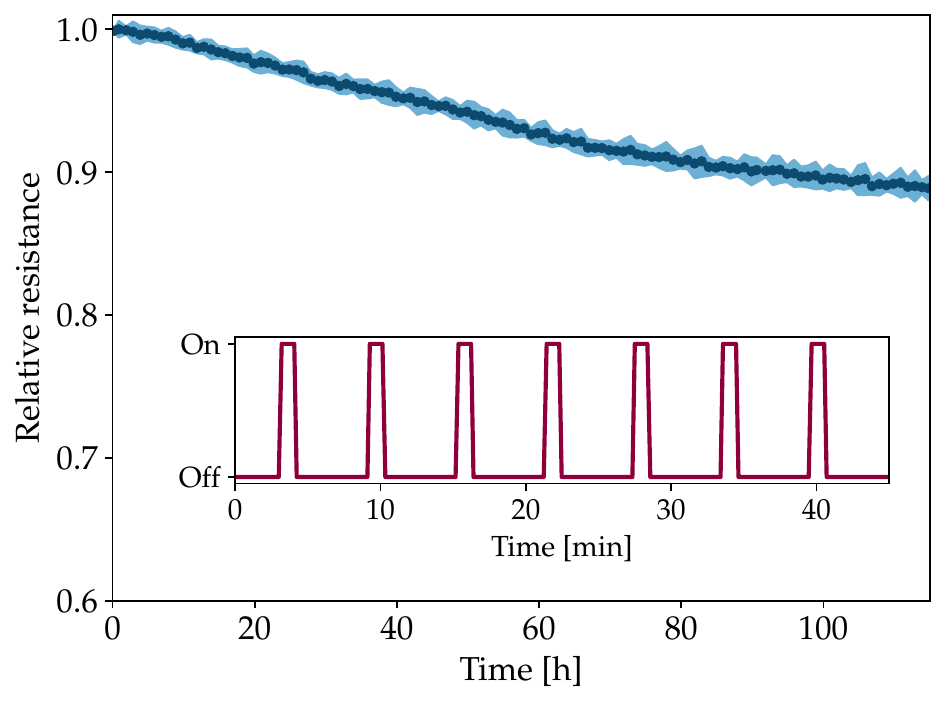"}
		\caption{\label{fig:resistance_vs_time} Relative resistance of the heating spiral during a repeated heating and cooling cycle. Dark blue points represent hourly averaged resistances and the light blue shaded area describes the corresponding standard deviation. Inset: Heating sequence shown for a fraction of the total measurement time. On and off refer to whether the electrical heating is active.}
	\end{figure}

	To investigate the long-term characteristics of the oven, we continuously operated it over the course of $11\,$days with MOT loading rates exclusively above $10^7\,\mathrm{atoms}/\mathrm{s}$. We observed no visible reduction of the size of the Yb block, suggesting a reasonably long lifetime of the oven at the given loading rate. 
	However, we observed an increase in the required optical heating power to maintain our target loading rate. We attribute this to an observed coating of the waveplate inside the pyramid from the evaporated ytterbium from the oven. Such a coating of the optical surfaces might be avoided by blocking the direct line of sight from the oven with a shutter, or by generating a collimated atomic beam emerging from the oven with a nozzle \cite{senaratne_effusive_2015} or by implementing a 2D-MOT. 
	Furthermore, we investigated the robustness of the oven during repeated temperature changes. We performed more than 1100 repetitions of heating the oven with $200\,$mW electrical power for $1\,$min followed by $5\,$min of cool-down time. We measured the oven resistance during the heating (Fig.~\ref{fig:resistance_vs_time}). While we observe a slow reduction of the observed resistance, a stable long-term operation can be achieved by stabilizing the electrical power instead of the current. We attribute the reduction of the resistance to either a slow heating of the peripheral infrastructure or an annealing effect of the thin Pt layer.
	
	\section{Conclusion}
	We presented a low-SWaP oven for alkaline-earth-like elements and characterized it by loading a MOT generated with a pyramid reflector. The defect-free microstructuring of fused silica using the LIDE process was a key technology to achieve the presented results. Starting from room temperature, the oven takes less than $5\,$minutes to heat up to a temperature that yields a large MOT loading rate. For MOT loading rates above $10^8\,\mathrm{atoms}/\mathrm{s}$, it requires less than $250\,$mW electrical heating power or less than $150\,$mW optical heating power. While electrical heating leads to a stable flux of atoms, the MOT loading rate measured with optical heating fluctuates due to power fluctuations of the heating beam.
	We demonstrated that a MOT of the most abundant isotope $^{174}$Yb as well as the clock-relevant isotope $^{171}$Yb can be loaded and that transfer to the second MOT stage is possible with a transfer efficiency above $90\,\%$. The oven thus constitutes a promising candidate to serve as an atom source for a future transportable optical lattice clock.
	Possible modifications of the oven design would allow to adjust key parameters. The heating time can be reduced by modifying the mounting springs for an increased heat conductance, at the cost of a higher power consumption. A trade-off between oven lifetime and heating time can be chosen by varying the size of the reservoir.
	
	\section{Acknowledgements}
	We thank J. Koch for fruitful discussions about the oven design and operation. We acknowledge funding from the joint project ``Innovative Vacuum Technology for Quantum Sensors'' (InnoVaQ) funded by the German Federal Ministry of Education and Research (BMBF) as part of the funding program ``quantum technologies – from basic research to market''. (Contract numbers: 13N15915 and 13N15916). We further acknowledge funding by the Deutsche Forschungsgemeinschaft (DFG, German Research Foundation) under Germany’s Excellence Strategy – EXC-2123 QuantumFrontiers – 390837967. 
	\bibliography{bibliography_oven_paper}
	
\end{document}